# Combining machine learning techniques, microanalyses and large geochemical datasets for tephrochronological studies in complex volcanic areas: new age constraints for the Pleistocene magmatism of Central Italy

*Maurizio Petrelli*[*], *Roberto Bizzarri, Daniele Morgavi, Angela Baldanza, Diego Perugini*

*Department of Physics and Geology, University of Perugia, Piazza Università, IT 06123 Perugia, Italy*

\* Corresponding author:

*Tel:* +39 075 585 2607

*Fax:* +39 075 585 2603

*e-mail:* maurizio.petrelli@unipg.it

*Abstract*

Characterization, correlation and provenance determination of tephra samples in sedimentary sections (tephrochronological studies) are powerful tools for establishing ages of depositional events, volcanic eruptions, and tephra dispersion. Despite the large literature and the advancements in this research field, the univocal attribution of tephra deposits to specific volcanic sources remains too often elusive. In this contribution, we test the application of a machine learning technique named Support Vector Machine to attempt shedding new light upon tephra deposits related to one of the most complex and debated volcanic regions on Earth: the Pliocene-Pleistocene magmatism in Italy. The machine learning algorithm was trained using one of the most comprehensive global petrological databases (GEOROC); 17 chemical elements including major ($SiO_2$, $TiO_2$, $Al_2O_3$, $Fe_2O_{3T}$, CaO, MgO, MnO, $Na_2O$, $K_2O$, $P_2O_5$) and selected trace (Sr, Ba, Rb, Zr, Nb, La, Ce) elements were chosen as input parameters. We first show the ability of support vector machines in discriminating among different Pliocene-Pleistocene volcanic provinces in Italy and then apply the same methodology to determine the volcanic source of tephra samples occurring in the Caio outcrop, an Early Pleistocene sedimentary section located in Central Italy. Our results show that: 1) support vector machines can successfully resolve high-dimensional tephrochronological problems overcoming the intrinsic limitation of two- and three-dimensional discrimination diagrams; 2) support vector machines can discriminate among different volcanic provinces in complex magmatic regions; 3) in the specific case study, support vector machines indicate that the most probable source for the investigated tephra samples is the so-called Roman Magmatic Province. These results have strong geochronological and geodynamical implications suggesting new age constraints (1.4 Ma instead of 0.8 Ma) for the starting of the volcanic activity in the Roman Magmatic Province.

## 1. INTRODUCTION

Machine Learning (ML) algorithms are a set of numerical tools capable of unraveling hidden structures in datasets to classify, predict or perform strategic decisions (Murphy, 2012). ML algorithms are generally divided into two main groups: (1) supervised and (2) unsupervised. Supervised algorithms (1) are trained using labeled data and, therefore, they learn by examples (Bishop, 2007). On the contrary, unsupervised algorithms (2) are trained by input vectors without any corresponding target value. As a consequence, the goal of unsupervised algorithms is to decipher hidden patterns without external suggestions. Example applications of supervised and unsupervised algorithms are the classification (i.e. assigning unknown samples to the correct class) and the clustering (i.e. discovering groups of similar examples within the datasets), respectively (Bishop, 2007). It is notable that the application of ML techniques has been quite extensively tested in the Earth Sciences (Abedi et al., 2012; Cannata et al., 2011; Goldstein and Coco, 2014; Huang et al., 2002; Masotti et al., 2006; Petrelli et al., 2003; Petrelli and Perugini, 2016; Zuo and Carranza, 2011) but their use is still virtually unexplored in tephra studies.

Among ML methods, Support Vector Machines (SVMs) consist of supervised learning algorithms particularly useful in the solution of problems where the goal is the correct classification of unknown samples (Cortes and Vapnik, 1995; Furey et al., 2000; W. Li et al., 2015; Noble, 2006; Petrelli and Perugini, 2016; Shi et al., 2015; Ustuner, 2015; Zuo and Carranza, 2011). It is significant to note that: 1) SVMs are particularly useful when dealing with high-dimensional datasets; 2) SVMs are widely tested on real-world problems; 3) SVM are still performing on data sets that have many attributes, even if there are very few cases on which to train the model (Cortes and Vapnik, 1995; Milenova et al., 2005). All these features make SVMs a potentially powerful tool for tephrochronological studies.

Tephrochronology (i.e. the study of volcanic products such as tephra layers, isolated glass fragments, and/or crystals embedded into sedimentary deposits) aids stratigraphers and volcanologists to perform long-range spatial and geochronological correlations (e.g. Bloom et al., 2012; Lowe, 2011; Satow et al., 2015). The recognition of compositionally similar volcanic tephra in layers from different sedimentary sequences provides a precise and well-established correlation tool, which is useful for establishing ages of depositional or volcanic events, tephra dispersion, and provenance of volcanic material (e.g. Bloom et al., 2012; Lowe, 2011; Satow et al., 2015).

As reported by Lowe (2011), geochemical correlations in tephra studies are often performed by plotting the supposed volcanic sources together with unknown tephra samples in binary and ternary discrimination diagrams. Unfortunately, recent studies (Li et al., 2015; Snow, 2006) highlighted that this approach, although particularly appealing because of its simplicity, might result in erroneous interpretations, especially when dealing with large datasets. As an example, Snow (2006) demonstrated that, due to visualization requirements, the success of binary and ternary discrimination diagrams is mainly hindered by their limited dimensionality.

In order to overcome these limitations, different multivariate approaches have been tested (Lowe, 2011 and references therein). Examples are the Principal Component Analysis (Pouget et al., 2014), the Discriminant Function Analysis (Lowe et al., 2007; Shane and Froggatt, 1994), the Manhalobis distance (Aksu et al., 2008), and the hierarchical cluster analysis (Preece et al., 2011).

In the present contribution we investigate, for the first time, the use of machine learning algorithms in high-dimensional tephrochronological studies and we combine them with a large, fully accessible, petrological databases: the GEOROC (GEOchemistry of Rocks of the Oceans and Continents) database. In detail, we first demonstrate the ability of SVMs in discriminating among the different volcanic provinces belonging to the complex Pliocene-Pleistocene magmatism in Italy (e.g. Peccerillo and Frezzotti, 2015, and references therein) using a large number of chemical elements (high-dimensional problem) as input parameters and a large number of samples to train the algorithm. Then we apply SVMs to attempt to recognize the volcanic source of tephra samples occurring in the Caio outcrop, an Early Pleistocene sedimentary section located in central Italy. We successively discuss the advantages and the limitations of the investigated methodology. We finally discuss the obtained results in the light of their implications for the evolution of the Pliocene-Pleistocene magmatism in central Italy and the potential impact on geodynamic interpretations.

## 2. MATERIALS AND METHODS

*2.1. Geochemical database for the Italian magmatism*

The reference dataset (i.e. the known samples utilized to train and test the model) derived from the GEOROC (http://georoc.mpch-mainz.gwdg.de/georoc/) database. This is one of the most comprehensive global petrological databases available at the moment and, therefore, it is suitable for the purposes of the present study. In detail, we extracted from the GEOROC database all volcanic samples characterized by a whole geochemical characterization of major ($SiO_2$, $TiO_2$, $Al_2O_3$, $Fe_2O_{3T}$, CaO, MgO, MnO, $Na_2O$, $K_2O$, $P_2O_5$) and selected trace (Sr, Ba, Rb, Zr, Nb, La, Ce) elements. The choice of the chemical elements to be analyzed by the system relies on the fact that they can be readily analyzed by X-ray fluorescence and inductively coupled plasma mass spectrometry for whole rocks samples and by common micro-analytical techniques (e.g. electron microprobe and laser ablation inductively coupled plasma mass spectrometry) for glasses (shards and melt inclusions). The result is that all these elements are generally included in geochemical analyses of single samples in petrological databases (e.g. GEOROC). This is essential to obtain a statistically representative number of samples (Petrelli and Perugini, 2016). Sample extracted from the GEOROC database have been then filtered as follows: a) samples marked as altered were removed; b) only whole rock data, glasses and melt inclusions analyses were considered; c) determinations characterized by closures (i.e. the sum of the total elements) below 94% were removed. To evaluate the power of system using the largest as possible reference dataset, we did not perform any additional filtering based for example, on the ages and/or the geochemical affinity (e.g. alkaline or sub-alkaline) of the studied samples. The volcanic provinces and/or single volcanic sources utilized in our study are defined following Peccerillo (2005). They are (Figure 1): Island of Pantelleria (PI; 218 samples), Etna Volcano (EV; 598 samples), Vesuvius Volcano (VV; 413 samples), Phlegrean Fields (PF; 853 samples), Aeolian Islands (AI; 930 samples), Mount Vulture (MV; 74 samples), Ernici and Roccamonfina Province (ERP; 176 samples), Intra-Apennine Volcanic Province (IAVP; 15 samples), Roman Magmatic Province (RMP; 409 samples) and Tuscan Magmatic Province (TMP; 140 samples). The resulting database consists of 3843 samples fully characterized for the considered 17 chemical elements (i.e. 17 dimensions). The ages and the main petrologic characteristics of the investigated magmatic provinces are reported in Table 1. As reported in Peccerillo (2005), the Tyrrhenian Sea region is one of the most complex geodynamic settings on Earth and its complexity is clearly expressed by the large variety of Plio-Quaternary volcanic rocks erupted in the area. It ranges from subalkaline to ultra-alkaline, from mafic to silicic, and from oversaturated to strongly undersaturated in silica. Trace element contents and isotopic ratios are also variable. Please refer to Peccerillo (2005) and Peccerillo and Frezzotti (2015) for further details.

*2.2. The case study site: Caio section*

The Early Pleistocene Caio sedimentary section is located in central Italy, about 20 km east of the town of Orvieto (Figure 1). The section (Figure 2) is ca. 30 m thick and consists of (1) a lower portion (ca. 23m) of shallow marine deposits and (2) an upper part (ca. 6m) characterized by continental deposits (Bizzarri et al., 2003). In the lower portion, three main facies associations are identified (Figure 2): (i) about 15 m of weakly cemented medium to fine sands, with wavy to channelled fine pebbles and/or bioclastic lags; (ii) about 3 m of silty sand/very fine sand, with bioclastic lags; (iii) about 5 m of grey, plain-parallel laminated silty clay, with gravel beds interposed in the last meter. Spotted molluscs (mainly oysters, pectinids and marine gastropods) widely occur. The final portions of the section are represented by mixed sand and gravel deposits of alluvial origin. Here, six sedimentary facies associations have been described (Figure 2): (a) lenticular shaped, clast-supported gravel (Gm) alternate to cross-laminated sand lenses (Sp); (b) parallel–laminated grey to green sandy clay (Sh); (c) matrix-supported, roughly organized fine to medium cobbles gravel (Gmc), with a(p)a(i) imbrications; (d) clast-supported, parallel cross-stratified very coarse pebbles gravel (Gp); (e) normally graded, medium to fine parallel cross-laminated sands (Sp); (f) silty-clayey, massive and very fine sands, with roots and concretions (paleosol). Further details are reported in Bizzarri et al. (2003). The facies analysis led to identifying, from the bottom to the top, a shoreface environment, a prodelta/fan-delta front environment, and an alluvial fan environment (Figure 2).

The fossil content and the biostratigraphy characterizing the lowermost meters, up to 19 m, document shallow water assemblages, dominated by molluscs and benthic foraminifera (Bizzarri et al., 2003). On other sections of the study area, such assemblages are associated with the Gelasian stage (*G. inflata* Zone: Baldanza et al., 2014, 2011; Bizzarri et al., 2015; Martinetto et al., 2015). On the interval between 19 m and 23 m, Calcareous Nannofossil assemblages of small *Gephyrocapsa* spp. (3.5-4 µm), medium *Gephyrocapsa* spp. (>4 µm), *Calcidiscus macintyrei*, *Helicosphaera sellii*, and *Coccolithus pelagicus* are found. These assemblages are referred to Gelasian *p.p.* - Calabrian (CNPL6 *p.p.*-CNPL7 Zones, CNPL8 Zone *sensu* Backman et al., 2012). Particularly, the First Occurrence (FO) of medium *Gephyrocapsa* spp. at 19 m, dated between 1.75 Ma and 1.71 Ma (Backman et al., 2012; de Kaenel et al., 1999; Di Stefano, 1998; Raffi, 2002; Rio et al., 1990), allow proposing this age for the lowermost volcanic episode (Figure 2). Thus, the two age limits for the lower section can be safely fixed to ~2.10-1.90 Ma and to ~1.60 Ma, respectively. According to various authors (Baldanza et al., 2011; Crippa et al., 2016) these age constraints are confirmed by the contemporaneous occurrence of *Globorotalia inflata*, *Hyalinea balthica*, and *Bulimina marginata* inside the clay deposits of facies association (iii). In addition, Late Villafranchian freshwater mollusc assemblages (Figure 2), collected in the upper deposits and dominated by *Theodoxus groyanus*, *Melanopsis affinis*, *Unio* cf. *U. pillai*, *Planorbis* sp. (Bizzarri et al., 2003), point to an age not younger than 1.4 Ma for the top of the section (Tasso-Farneta F.U.: Gliozzi et al., 1997). As a consequence, the age of the upper portion of Caio section can be safely fixed between 1.71 Ma and 1.4 Ma.

Volcanic deposits are widely represented throughout the Caio section (Figure 2); in particular, two intervals ranging from 18 m to 19.5 m and from 22.5 m to 28.5 m, respectively, are enriched in small fragments of pyroclastic materials, mainly pumices and isolated clinopyroxenes.

Pyroxenes and pumices increase in abundance from the bottom to the top of the section. Pumices are angular to sub-angular, with minimal evidence of transport; most pyroxene crystals are idiomorphic with little or no evidence of sedimentary reworking (Figure 3a) indicating a deposition of the volcanic material directly from the pyroclastic clouds into the sedimentary basin and/or a sedimentary reworking near contemporaneous to the primary event.

*2.3. Sample collection, preparation and optical investigations*

Five samples, ca. 3 kg in weight each, were collected from the Caio section at different heights in the stratigraphic section in order to obtain a complete characterization of the upper portion of Caio deposits (Figure 2). Pyroxenes and pumices were separated from epiclastic material by hand picking. Pumice dimensions range between 2 and 5 mm whereas pyroxenes range between 0.5 and 5 mm.

Petrographic observations on pumice samples indicated a porphyritic texture with microphenocrysts of idiomorphic clinopyroxene, leucite, and oxides embedded in a vesicular, glassy to variably microcrystalline groundmass (Figure 3b). Leucite appears extensively transformed to analcime. The groundmass ranges from glassy to microcrystalline, often showing secondary oxidation. Calcite crystals occasionally occur as secondary phases filling voids. Due to the extensive secondary alteration occurring in pumice samples, pumices were not used for chemical analysis.

Detailed observation of pyroxene crystals (Figure 3c and 3d) clearly revealed the presence of alteration-free melt-inclusions ranging in size from a few μm to about 200 μm to be potentially investigated for major and trace elements by micro-analytical determinations. Isolated pyroxene crystals have been therefore washed, cast in epoxy resin, prepared as 60 μm thick thin sections and finally polished up to 1 μm using progressively finer diamond pastes. Analysed melt inclusions were carefully selected to avoid those inclusions showing crystallization, devitrification and decrepitation phenomena.

*2.4. Laser Ablation-Inductively Coupled Plasma-Mass Spectrometry*

Laser Ablation-Inductively Coupled Plasma-Mass Spectrometry (LA-ICP-MS) analyses were performed at the Department of Physics and Geology, Perugia University. The LA-ICP-MS instrumentation consists of a commercial New Wave UP213 (ESI - New Wave, UK) LA system coupled with a Thermo X Series (Thermo Fisher Scientific, Bremen, DE) ICP-MS. The laser ablation system is a frequency quintupled Nd:YAG laser, whose fundamental wavelength of 1064 nm is converted into 213 nm using three harmonic generators.

Helium was preferred over argon as a carrier gas to enhance the transport efficiency of the ablated aerosol (Eggins et al., 1998). The helium carrier gas exiting the ablation cell was mixed with argon make-up gas before entering the ICP torch; this configuration allowed for the maintenance of stable and optimum excitation condition (Alagna et al., 2008; Petrelli et al., 2016, 2008, 2007).

The LA-ICP-MS system was optimized for dry plasma conditions before each analytical session on a continuous linear ablation of NIST SRM 612 glass standard by maximizing the signals for selected masses ($La^+$ and $Th^+$) and reducing oxide formation by minimizing the $ThO^+/Th^+$ ratio. The U/Th ratio was also monitored to be close to 1. Only exposed melt inclusions and larger than 50 μm were analysed using a laser beam diameter of 30 μm.

To obtain a complete geochemical characterization of the studied samples, all major and trace elements were recorded in the same LA-ICP-MS run following the protocol proposed by Guillong et al. (2005) for major elements and Longerich et al. (1996) for trace elements.

Major elements (SiO$_2$, TiO$_2$, Al$_2$O$_3$, Fe$_2$O$_{3T}$, CaO, MgO, MnO, Na$_2$O, K$_2$O, P$_2$O$_5$) data reduction was performed using the USGS BCR2G (Rocholl, 1998) as calibrator and normalizing the results to a fixed oxide total following the procedure described in Guillong et al. (2005). The concentration of one major element ($^{29}$Si) was subsequently used as internal standard to reduce trace element concentrations from count per second signals following the method proposed by Longerich et al. (1996). External calibration for trace elements was performed using the NIST SRM-612 (Pearce et al., 1997) reference material. For a complete description of the LA-ICP-MS instrumentation used in this work please refer to (Petrelli et al., 2008). Host minerals were always analysed close to the studied melt inclusions to investigate potential chemical exchanges between the trapped melt and the mineral.

*2.5. Support Vector Machines*

In SVMs, input vectors are mapped to a very high-dimension feature space. In this feature space a decision surface is then constructed (Cortes and Vapnik, 1995). In the simplest implementation, a support vector machine constructs a hyper-plane or set of hyper-planes in a high-dimensional space, which can be used for classification, regression or other tasks. Intuitively, a good separation is achieved when the hyper-plane has the largest distance to the nearest training data points of any class (so-called functional margin), since in general the larger the margin the lower the generalization error of the classifier (Cortes and Vapnik, 1995). The simplest way to separate two groups of data is through a straight line (2 dimensions, in our case two chemical species as in the case of binary classification diagrams), a flat plane (3 dimensions, three chemical elements) or an N-dimensional hyper-plane (i.e. N chemical elements). However, there are cases where a non-linear region can separate the groups more efficiently. SVMs handle these occurrences using non-linear kernel functions. This means that a non-linear function is learned as a linear problem in a high-dimensional feature space. This is called "kernel trick" and means that the kernel function transforms the data into a higher dimensional feature space to allow for a linear separation (Cortes and Vapnik, 1995).

The mathematical formulation of SVM problems can be defined as follow. Consider a training dataset of S-dimensional samples (e.g. S chemical elements as input) $x_i$ with $i=1,2,3,\ldots,n$ where $n$ is the number of samples. For a two-class problem also consider a vector $y \in \{1, -1\}^n$. SCV solves the following primal problem:

$$\max_{\alpha} \left[ \sum_{i=1}^{n} \alpha_i - \frac{1}{2} \boldsymbol{\alpha}^T \boldsymbol{H} \boldsymbol{\alpha} \right] \quad s.t. \quad 0 \leq \alpha_i \leq C \; \forall_i \quad and \quad \sum_{i=1}^{n} \alpha_i y_i = 0$$

where H is an $n$ by $n$ positive semidefinite matrix $H_{i,j} \equiv y_i y_j K(x_i x_j)$ where $K(x_i x_j)$ is the kernel function. To solve the multiclass problem, the One Vs One (OVO) approach has been utilized (Hsu and Lin, 2002). In OVO, one SVM classifier is built for all possible pairs of classes (Dorffner et al., 2001; Knerr et al., 1990). The output from each classifier is obtained in the form of a class label. The class label with the highest frequency is assigned to that point in the data vector (Hsu and Lin, 2002). The reader interested in the details of the SVM theory and numerical methods can find full methodological descriptions in (Cortes and Vapnik, 1995).

*2.6. SVMs Implementation*

To evaluate the performance of SVMs in tephrochronological applications, the scikit-learn (Pedregosa et al., 2011), machine learning Python module has been used. The use of a general-purpose, high-level, language (Pedregosa et al., 2011) is motivated by the fact that it allows for the use of complex machine learning techniques even to scientists not deeply involved in the field of artificial intelligence. All the Python scripts utilized in the present study, together with the complete dataset and instructions to setup a Python environment suited for ML applications are available at the following public web site version control system repository (Anderson and Smith, 1995): https://goo.gl/Hd5x7o.

All major (SiO$_2$, TiO$_2$, Al$_2$O$_3$, Fe$_2$O$_{3T}$, CaO, MgO, MnO, Na$_2$O, K$_2$O, P$_2$O$_5$) and selected trace (Sr, Ba, Rb, Zr, Nb, La, Ce) elements were utilized as input parameter. The relative role of major and trace elements in a SVM classification have been discussed extensively in (Petrelli and Perugini, 2016). In general, the combination of major and trace elements always shows better performance relative to the separate use of these two groups of elements. This is due to the fact that the larger the number of variables to be analyses, the higher the probability to associate a tephra sample to the correct volcanic source (see Petrelli and Perugini, 2016 for further details).

We opted for the Radial Basis Function (RBF) as kernel function:

$$K(x_i, x_j) = e^{-\gamma(x_i - x_j)^2}$$

This choice is motivated by the fact that it is one of the most widely used and performing non-linear kernels (Scholkopf et al., 1997). Hence, using an SVM with an RBF learning kernel, the system depends on two tunable parameters: $\gamma$ and $C$. $C$ is the penalty parameter of the error term. It trades off misclassification of training examples against the simplicity of the decision surface. The $\gamma$ parameter can be seen as the inverse of the radius of influence of samples selected by the model as support vectors.

*2.7. Data standardization and learning metrics*

Standardization of a dataset is a common requirement for many machine-learning estimators. As an example, the RBF function utilized in the present study is one of the estimators requiring a standard, normally distributed dataset.

In detail, we first arranged all the analysis to 100% on an anhydrous basis (e.g. Verma and Armstrong-Altrin, 2016). As a second step, we applied the box-cox transformation (Box and Cox, 1964) to obtain near-normal distributions . Finally, the obtained distributions are average subtracted and normalized to a variance equal to 1 to overcome the biases resulting to the simultaneous statistical treatment of major and trace elements (e.g. Templ et al., 2008). The potential bias related to so-called 'closure effect' (Aitchison, 1992; Buccianti et al., 2006; Butler, 1976) have been also investigated by opening the system using the isometric log-ratio transformation (Egozcue et al., 2003) and comparing results obtained using the box-cox transformation. Results do not show appreciable differences. The outcomes from two procedures can be evaluated using the Python script files provided on line (https://goo.gl/Hd5x7o).

In order to have a metric providing information about the ability of the system to classify correctly unknown samples to the relative volcanic source, we define a quantity, which we name Classification Score (*CS*). In detail, *CS* (%) is quantitatively defined as the ratio of the number of correctly classified (*CC*) samples and the total number of samples (*n*) constituting the studied population:

$$CS\ [\%] = 100 \cdot \frac{CC}{n}$$

*2.8. Experiments*

To evaluate the learning capabilities of SVMs when dealing with a tephrochronological application, we split the whole dataset into two randomly distributed sub-populations containing the 70% (learning population) and 30% (test population) of samples, respectively. Then, the learning population was utilized to train the ML algorithms and the test population was analyzed as unknown.

Since the learning capabilities of the system depend on *C* and $\gamma$ parameters (see section 2.5), in the first experimental set we evaluated the classification scores on a two-dimensional 13x13 grid of *C* and $\gamma$ values ranging from $10^{-2}$ to $10^{10}$ and from $10^{-8}$ to $10^{4}$, respectively. This procedure, also known as algorithm tuning, allowed us to define the best *C* and $\gamma$ values to be used in the present study.

As a second step, we utilized the two *C* and $\gamma$ values characterized by the highest classification scores to investigate the discrimination power of the SVM for the different Italian magmatic provinces using a Leave One Out approach (LOO) (James et al., 2013). LOO is one of the simplest cross-validation methods and consists in learning the system by taking all the samples except one. The sample which is left out is then introduced into the system as unknown. Thus, for $n$ samples, we performed $n$ different trainings, one for each unknown sample.

Finally (third experimental set), we applied SVMs in the attempt to solve the case study of tephra products from the Caio section to determine the potential volcanic source.

## 3. RESULTS AND DISCUSSION

Figure 4 shows the results of the tuning for the SVM algorithm plotted on a 13x13 grid of $C$ and $\gamma$ values. Dark colours in the figure correspond to low classification scores (close to 20 %) whereas light colours highlight classification scores approaching 100%. Figure 4 shows that the highest classification score ($CS$ = 98 %) was obtained for $C$ and $\gamma$ values of 100 and 0.1, respectively.

Therefore, the obtained $C$ and $\gamma$ values have been utilized to evaluate the potential of the SVM in discriminating among the different volcanic sources that were active during the development of Pliocene and Pleistocene Italian magmatism. Results are reported in Figure 5 in the form of a so-called "confusion matrix" (Provost and Kohavi, 1998). In a confusion matrix, each column represents the instances in a predicted class, whereas each row represents the instances in an actual class. As a consequence, correct estimations are reported in the cells belonging to the main diagonal of the matrix whereas the errors are reported in the other cells. Observing Figure 5, it emerges the picture that all the investigated volcanic sources are characterized by $CS$ values above 90 %. In particular, obtained $CS$ values are 98.1 % for the Vesuvius Volcano (VV), 99.5 % for the Etna Volcano (EV), 98.1 % for the Phlegrean Fields area (PF), 98.7 % for the Aeolian Island archipelago (AI), 97.6 % for the Roman Magmatic Province (RMP), 96.4 % for the Tuscan Magmatic Province (TMP), 98.6 % for Pantelleria Island (PI), 93.3% for the Intra Apennine

Volcanic Province (IAVP), 98.6 % for Monte Vulture (MV) and 90.9 % for the Ernici-Roccamonfina Province (ERP).

As the final step of our analysis, we applied our model to the 17 analyzed samples of to the Caio section (Table 2). Among the 17 analyzed samples, 14 are assigned to the Roman Magmatic Province (RMP) and three to the Ernici-Roccamonfina Province (ERP). These results clearly indicate that the RMP is the likely source for the tephra products embedded in the Caio Section. Further support to this hypothesis is provided by qualitative geochemical observations indicating that: 1) the investigated samples are of ultrapotassic geochemical affinity; 2) analysed samples are undersaturated in silica; 3) rare earth elements patterns are fractionated with LREE progressively enriched relative to HREE (Figure 6A); 4) the patterns of mantle normalized incompatible elements are characterized by high LILE enrichments and high LILE/HFSE patterns (Figure 6B). The fact that geochemical observations point qualitatively to the same conclusions as those derived from the application of the ML approach argues in favor of the robustness and reliability of this method. It is to note, however, that geochemical observations only provide qualitative information and, from this point of view, they are rather subjective. On the contrary, the ML approach provides quantitative information and is statistically robust.

It is notable at this point that the use of binary discrimination diagrams is unable to univocally identify the RMP as a source for Caio tephra. As an example, Figures 7 and 8 report a Total Alkali vs. $SiO_2$ and a Zr vs. Nb diagrams respectively where Caio tephra are plotted together with all the samples belonging to the different magmatic provinces (drawn as areas) considered in our study. From the observation of Figure 7, it emerges that major elements alone might be able to to exclude only some of the investigated magmatic provinces. These are the TMP, EV, PF, PI and AI. The fields defined by samples belonging to these magmatic provinces do not overlap with Caio samples and, in addition, these samples miss one or more of the main petrological features of Caio tephra (e.g. ultrapotassic geochemical affinity or silica undersaturation). Nonetheless, there is still a considerable overlapping with other magmatic provinces (RMP, ERP and IAVP, MV and VV) that cannot be discriminated. Trace element binary diagrams (Figure 8) do not improve the discrimination for Caio samples and continue to show a large overlap with at least three of the remnant potential magmatic sources: RMP, ERP and IAVP. Other discriminating diagrams (such as Ba/La vs. Zr/Nb and Th/Ta vs. Ce/Sr; Peccerillo and Frezzotti, 2015) display similar results. Therefore, although Caio tephra samples always overlap with compositions from RMP, they are not easily discriminable from the other provinces using classical discriminant diagrams, denying the possibility to attribute these samples to a specific volcanic source.

Up to now the RMP was considered active between 0.8 and 0.02 Ma (e.g. Peccerillo and Frezzotti, 2015). The area was interpreted as a series of volcanic complexes constituted by the Monti Vulsini, Vico, Monti Sabatini and Colli Albani (Alban Hills) districts. The products are superimposed to the igneous rocks of the Tuscany magmatic to the north, and there is petrological evidence for interaction between the RMP and the Tuscany province (Conticelli et al., 2007; Conticelli and Peccerillo, 1992).

The results of our study clearly indicate the occurrence of volcanic products, analyzed here for the first time, that are older then the RMP province, but still characterized by an unambiguous RMP geochemical affinity. This suggests that volcanic activity in the RMP should be shifted to at least 1.4. Ma ago, i.e. more than 0.6 Ma before the age considered up to now (Marra et al., 2014; Peccerillo and Frezzotti, 2015) for the starting of magmatism in the RMP. We recognize that our results are based on the occurrence of tephra samples on a single sedimentary outcrop and that they must be supported by further studies based on tephra occurring on a statistically representative number of sedimentary sections before starting the re-dating of the magmatism associated to the RMP.

If confirmed, our results will have also strong magmatological and geodynamical implications such they clearly indicate a larger overlapping in time between the RMP and the TMP. This might have a strong impact for the time development of TMP and RMP in terms of source regions of magmas as well as for the timing of metasomatic processes acting in the mantle wedge below the study area and the relative geodynamic implications. We believe that results presented here, therefore, will fuel novel interpretations of this important crustal sector that will be the focus of future investigations.

Furthermore, it is significant to note that the Caio section in which the analyzed tephra were found is not the only place in which analogous volcanic products are reported in the area. Baldanza et al. (2011) and Bizzarri et al. (2015) documented the presence of a 20 km wide area of volcanic products around the town of Orvieto, both in continental and/or marine sections bio-stratigraphically dated to Early Pleistocene. These volcanic products are reported, in recent stratigraphic schemes, as "pre Vulsini volcanic events", inside the "Chiani-Tevere" informal Lithostratigraphic Unit (Baldanza et al., 2014; Martinetto et al., 2015). Their study might also provide further constraints for understanding the timing of volcanic events of this geologic area and the relative geodynamic implications.

## 4. Summary and Concluding Remarks

In this work we have shown how the combined use of ML techniques and large geochemical datasets can be successfully applied to solve tephrochronological and tephrostratigraphical problems. In detail, our results indicate that: 1) support vector machine can resolve high-dimensional tephrochronological problems; 2) support vector machines can discriminate among the different volcanic provinces even in complex geologic settings such as the Pliocene-Pleistocene volcanic areas of Italy; 3) support vector machines unambiguously indicate a Roman geochemical affinity for the pyroclastic products embedded in the Caio stratigraphic section; 4) the age of the starting activity of the Roman Magmatic Province must be reconsidered in the light of our results and has to be likely shifted back to at least 1.4 Ma. The presented results are likely to open new perspectives for the interpretation of the evolution of the Pliocene magmatism in central Italy.

Although ML methods can be successfully applied and might potentially open new frontiers in tephra studies, it may be worth discussing their limitations. They can be summarized as follows (Petrelli and Perugini (2016)): 1) ML techniques must not be considered as a "black box" where input geochemical data are "magically" transformed into classification diagrams to be immediately utilized for scientific interpretations; 2) ML methods require the integration with other techniques such as fieldwork, petrographic observations and classic geochemical studies to obtain a clearer picture of the investigated problem; 3) the proposed ML system fails if the right volcanic source is not present among the ones utilized for the learning stage.

In the specific case of recognition of volcanic source starting from tephra layers, additional issues can arise from the heterogeneous nature of the GEOROC database where (a) bulk (e.g. XRF and INAA) and micro-analytical (e.g. EPMA and LA-ICP-MS) determinations are mixed and (b) it is not easy to extract information at the level of single eruptions. Published studies highlighted that bulk geochemical data derived by micro-analytical techniques compare well with bulk rock analyses (Pearce et al., 1999; Schmid et al., 2000; Tomlinson et al., 2015) indicating that the mixed nature of geochemical databases is unlikely to be a major problem. As for the point (b), several recent large scale projects (e.g. Blockley et al. 2014; Bronk Ramsey et al., 2015; Hoek et al., 2008; Lowe et al., 2015) started filling the gap with the generation of detailed and well constrained datasets at the level of single eruptions opening new perspectives for the correlation of tephra layers to specific eruptions. As a consequence, the next future challenge might be the application of ML techniques to unravel and classify single eruptions using more specific and detailed source databases.

**Acknowledgements**

We wish to acknowledge the European Research Council for the Consolidator Grant CHRONOS (no. 612776), the Department of Physics and Geology, University of Perugia for the "CHALLENGE" FRB 2015 grant (Maurizio Petrelli, PI) and the Microsoft Research Azure Award Program (CRM:0518576 - Maurizio Petrelli).

**References**

Abedi, M., Norouzi, G.-H., Bahroudi, A., 2012. Support vector machine for multi-classification of mineral prospectivity areas. Comput. Geosci. 46, 272–283. doi:10.1016/j.cageo.2011.12.014

Aitchison, J., 1992. On criteria for measures of compositional difference 24, 365–379. doi:10.1007/BF00891269

Aksu, A.E., Jenner, G., Hiscott, R.N., İşler, E.B., 2008. Occurrence, stratigraphy and geochemistry of Late Quaternary tephra layers in the Aegean Sea and the Marmara Sea. Mar. Geol. 252, 174–192. doi:10.1016/j.margeo.2008.04.004

Alagna, K.E., Petrelli, M., Perugini, D., Poli, G., 2008. Micro-analytical zircon and monazite U-Pb isotope dating by laser ablation-inductively coupled plasma-quadrupole mass spectrometry. Geostand. Geoanalytical Res. 32, 103–120. doi:10.1111/j.1751-908X.2008.00866.x

Anderson, J.L., Smith, D.R., 1995. The effects of temperature and fO2 on the Al-in-hornblende barometer. Am. Mineral.

Backman, J., Raffi, I., Rio, D., Fornaciari, E., Pälike, H., 2012. Biozonation and biochronology of Miocene through Pleistocene calcareous nannofossils from low and middle latitudes. Newsl. Stratigr. 45, 221–244.

Baldanza, A., Bizzarri, R., Famiani, F., Pasini, G., Garassino, A., De Angeli, A., 2014. Early Pleistocene shallow marine decapod crustaceans fauna from Fabro Scalo (western Umbria, central Italy): taxonomic inferences and palaeoenvironmental reconstruction. Neues Jahrb. für Geol. und Paläontologie - Abhandlungen 271, 261–283. doi:10.1127/0077-7749/2014/0389

Baldanza, A., Bizzarri, R., Hepach, H., 2011. New biostratigraphic data from the Early Pleistocene tyrrhenian PALEOCOAST (Western Umbria, Central Italy). Geol. Croat. 64, 133–142. doi:10.4154/gc.2011.11

Bishop, C., 2007. Pattern recognition and machine learning. Springer Verlag, New York.

Bizzarri, R., Ambrosetti, P., Argenti, P., Gatta, G.D., Baldanza, A., 2003. L'affioramento del Caio (Lago di Corbara, Orvieto, Italia centrale) nell'ambito dell'evoluzione paleogeografica Plio –

Pleistocenica della Valle del Tevere: evidenze sedimentologiche e stratigrafiche. Quat. Ital. J. Quat. Sci. 16, 241–255.

Bizzarri, R., Rosso, A., Famiani, F., Baldanza, A., 2015. Lunulite bryozoans from Early Pleistocene deposits of SW Umbria (Italy): sedimentological and paleoecological inferences. Facies.

Blockley, S.P.E., Bourne, A.J., Davies, S.M., Hardiman, M., Harding, P.R., Lane, C.S., MacLeod, A., Matthews, I.P., Pyne-O'Donnell, S.D.F., Rasmussen, S.O., Wulf, S., Zanchetta, G., 2014. Tephrochronology and the extended intimate (integration of ice-core, marine and terrestrial records) event stratigraphy 8–128 ka b2k. Quat. Sci. Rev. 106, 88–100. doi:10.1016/j.quascirev.2014.11.002

Bloom, J.S., Richards, J.W., Nugent, P.E., Quimby, R.M., Kasliwal, M.M., Starr, D.L., Poznanski, D., Ofek, E.O., Cenko, S.B., Butler, N.R., Kulkarni, S.R., Gal-Yam, A., Law, N., 2012. Automating Discovery and Classification of Transients and Variable Stars in the Synoptic Survey Era. Publ. Astron. Soc. Pacific 124, 1175–1196. doi:10.1086/668468

Box, G.E.P., Cox, D.R., 1964. An analysis of transformations. J. R. Stat. Soc. Ser. B 26, 211–252. doi:10.2307/2287791

Bronk Ramsey, C., Lane, C.S., Smith, V.C., Pollard, A.M., 2015. The RESET tephra database and associated analytical tools. Quat. Sci. Rev. 118, 33–47. doi:10.1016/j.quascirev.2014.11.008

Buccianti, A., Mateu-Figueras, G., Pawlowsky-Glahn, V., 2006. Compositional data analysis in the geosciences: from theory to practice. Geol. Soc. Lond. Spec. Publ.

Butler, J.C., 1976. Principal components analysis using the hypothetical closed array 8, 25–36. doi:10.1007/BF01039682

Cannata, A., Montalto, P., Aliotta, M., Cassisi, C., Pulvirenti, A., Privitera, E., Patanè, D., 2011. Clustering and classification of infrasonic events at Mount Etna using pattern recognition techniques. Geophys. J. Int. 185, 253–264. doi:10.1111/j.1365-246X.2011.04951.x

Cohen, K.M., Gibbard, P., 2011. Global chronostratigraphical correlation table for the last 2.7 million years. Subcommission on Quaternary Stratigraphy (International Commission on Stratigraphy), Cambridge, England.

Conticelli, S., Carlson, R.W., Widom, E., Serri, G., 2007. Chemical and isotopic composition (Os, Pb, Nd, and Sr) of Neogene to Quaternary calc-alkalic, shoshonitic, and ultrapotassic mafic rocks from the Italian peninsula: Inferences on the nature of their mantle sources, in: Special Paper of the Geological Society of America. Geological Society of America, pp. 171–202. doi:10.1130/978-0-8137-2418-8

Conticelli, S., Peccerillo, A., 1992. Petrology and geochemistry of potassic and ultrapotassic volcanism in central Italy: petrogenesis and inferences on the evolution of the mantle sources. Lithos 28, 221–240. doi:10.1016/0024-4937(92)90008-M

Cortes, C., Vapnik, V., 1995. Support-vector networks. Mach. Learn. 20, 273–297. doi:10.1007/BF00994018

Crippa, G., Angiolini, L., Bottini, C., Erba, E., Felletti, F., Frigerio, C., Hennissen, J.A.I., Leng, M.J., Petrizzo, M.R., Raffi, I., Raineri, G., Stephenson, M.H., 2016. Seasonality fluctuations recorded in fossil bivalves during the early Pleistocene: Implications for climate change. Palaeogeogr. Palaeoclimatol. Palaeoecol. 446, 234–251. doi:10.1016/j.palaeo.2016.01.029

de Kaenel, E., Siesser, W.G., Murat, A., 1999. Pleistocene calcareous nannofossil biostratigraphy and the western mediterranean sapropels, sites 974 to 977 and 979, in: Proceedings of the Ocean Drilling Program: Scientific Results. pp. 159–183.

Di Stefano, E., 1998. Calcareous nannofossil quantitative biostratigraphy of Holes 969E and 963B (eastern Mediterranean). Proc. Ocean Drill. Progr. Sci. Results 160, 99–112.

Dorffner, G., Bischof, H., Hornik, K., 2001. Artificial Neural Networks - ICANN 2001. Springer-Verlag, Berlin, Heidelberg.

Eggins, S.M., Kinsley, L.P.J., Shelley, J.M.G., 1998. Deposition and element fractionation processes during atmospheric pressure laser sampling for analysis by ICP-MS. Appl. Surf. Sci. 127–129, 278–286.

Egozcue, J.J., Pawlowsky-Glahn, V., Mateu-Figueras, G., Barceló-Vidal, C., 2003. Isometric

Logratio Transformations for Compositional Data Analysis. Math. Geol. 35, 279–300. doi:10.1023/A:1023818214614

Furey, T.S.T.T.S., Cristianini, N., Duffy, N., Bednarski, D.W., Schummer, M., Haussler, D., 2000. Support vector machine classification and validation of cancer tissue samples using microarray expression data. … 16, 906–914. doi:10.1093/bioinformatics/16.10.906

Gliozzi, E., Abbazzi, L., Argenti, P., Azzaroli, A., Caloi, L., Capasso Barbato, L., Di Stefano, G., Esu, D., Ficcarelli, G., Girotti, O., Kotsakis, T., Masini, F., Mazza, P., Mezzabotta, C., Palombo, M.R., Petronio, C., Rook, L., Sala, B., Sardella, R., Zanalda, E., Torre, D., 1997. Biochronology of selected mammals, molluscs and ostracods from the middle pliocene to the late pleistocene in Italy. The state of the art. Riv. Ital. di Paleontol. e Stratigr. 103, 369–388.

Goldstein, E.B., Coco, G., 2014. A machine learning approach for the prediction of settling velocity. Water Resour. Res. 50, 3595–3601. doi:10.1002/2013WR015116

Guillong, M., Hametner, K., Reusser, E., Wilson, S.A., Günther, D., 2005. Preliminary characterisation of new glass reference materials (GSA-1G, GSC-1G, GSD-1G and GSE-1G) by laser ablation-inductively coupled plasma-mass spectrometry using 193 nm, 213 nm and 266 nm wavelengths. Geostand. Geoanalytical Res. 29, 315–331.

Hoek, W.Z., Yu, Z.C., Lowe, J.J., 2008. INTegration of Ice-core, MArine, and TErrestrial records (INTIMATE): refining the record of the Last Glacial–Interglacial Transition. Quat. Sci. Rev. 27, 1–5. doi:10.1016/j.quascirev.2007.11.020

Hsu, C.-W.C., Lin, C.C.-J., 2002. A comparison of methods for multiclass support vector machines. Neural Networks, IEEE Trans. 13, 415–25. doi:10.1109/72.991427

Huang, C., Davis, L.S., Townshend, J.R.G., 2002. An assessment of support vector machines for land cover classification. Int. J. Remote Sens. 23, 725–749. doi:10.1080/01431160110040323

Iaccarino, S., Premoli Silva, I., 2007. Practical manual of Neogene planktonic foraminifera. – International School on Planktonic Foraminifera, VI Course: Neogene. Perugia.

James, G., Witten, D., Hastie, T., Tibshirani, R., 2013. An introduction to statistical learning.

Springer Verlag, New York.

Knerr, S., Personnaz, L., Dreyfus, G., 1990. Single-layer learning revisited: A stepwise procedure for building and training a neural network, in: Neurocomputing: Algorithms, Architectures and Applications. Springer International Publishing AG, pp. 41–50. doi:10.1007/978-3-642-76153-9_5

Li, C., Arndt, N.T., Tang, Q., Ripley, E.M., 2015. Trace element indiscrimination diagrams. Lithos 232, 76–83. doi:10.1016/j.lithos.2015.06.022

Li, W., Huang, H., Peng, F., 2015. Trajectory classification in circular restricted three-body problem using support vector machine. Adv. Sp. Res. 56, 273–280. doi:10.1016/j.asr.2015.04.017

Longerich, H.P., Jackson, S.E., Günther, D., 1996. Laser ablation inductively coupled plasma mass spectrometric transient signal data acquisition and analyte concentration calculation. J. Anal. At. Spectrom. 11, 899–904.

Lowe, D.J., 2011. Tephrochronology and its application: A review. Quat. Geochronol. 6, 107–153. doi:10.1016/j.quageo.2010.08.003

Lowe, J.J., Ramsey, C.B., Housley, R.A., Lane, C.S., Tomlinson, E.L., 2015. The RESET project: constructing a European tephra lattice for refined synchronisation of environmental and archaeological events during the last c. 100 ka. Quat. Sci. Rev. 118, 1–17. doi:10.1016/j.quascirev.2015.04.006

Lowe, J.J.J., Blockley, S., Trincardi, F., Asioli, A., Cattaneo, A., Matthews, I.P.P., Pollard, M., Wulf, S., 2007. Age modelling of late Quaternary marine sequences in the Adriatic: Towards improved precision and accuracy using volcanic event stratigraphy. Cont. Shelf Res. 27, 560–582. doi:10.1016/j.csr.2005.12.017

Marra, F., Sottili, G., Gaeta, M., Giaccio, B., Jicha, B., Masotta, M., Palladino, D.M., Deocampo, D.M., 2014. Major explosive activity in the Monti Sabatini Volcanic District (central Italy) over the 800–390 ka interval: geochronological–geochemical overview and tephrostratigraphic

implications. Quat. Sci. Rev. 94, 74–101. doi:10.1016/j.quascirev.2014.04.010

Martinetto, E., Momohara, A., Bizzarri, R., Baldanza, A., Delfino, M., Esu, D., Sardella, R., 2015. Late persistence and deterministic extinction of "humid thermophilous plant taxa of East Asian affinity" (HUTEA) in southern Europe. Palaeogeogr. Palaeoclimatol. Palaeoecol. doi:10.1016/j.palaeo.2015.08.015

Masotti, M., Falsaperla, S., Langer, H., Spampinato, S., Campanini, R., 2006. Application of Support Vector Machine to the classification of volcanic tremor at Etna, Italy. Geophys. Res. Lett. 33, L20304. doi:10.1029/2006GL027441

McDonough, W.F., Sun, S. -s., 1995. The composition of the Earth. Chem. Geol. 120, 223–253. doi:10.1016/0009-2541(94)00140-4

Milenova, B.L., Yarmus, J.S., Campos, M.M., 2005. SVM in oracle database 10g: removing the barriers to widespread adoption of support vector machines 1152–1163.

Murphy, K.P., 2012. Machine Learning: A Probabilistic Perspective. The MIT Press, Cambridge, MA.

Nakamura, N., 1974. Determination of REE, Ba, Fe, Mg, Na and K in carbonaceous and ordinary chondrites. Geochim. Cosmochim. Acta 38, 757–775. doi:10.1016/0016-7037(74)90149-5

Noble, W.W.S., 2006. What is a support vector machine? Nat. Biotechnol. 24, 1565–1567. doi:10.1038/nbt1206-1565

Pearce, N.J.G., Westgate, J.A., Perkins, W.T., Eastwood, W.J., Shane, P., 1999. The application of laser ablation ICP-MS to the analysis of volcanic glass shards from tephra deposits: bulk glass and single shard analysis. Glob. Planet. Change 21, 151–171. doi:10.1016/S0921-8181(99)00012-0

Pearce, N.J.G.G., Perkins, W.T., Westgate, J.A., Gorton, M.P., Jackson, S.E., Neal, C.R., Chenery, S.P., 1997. A Compilation of New and Published Major and Trace Element Data for NIST SRM 610 and NIST SRM 612 Glass Reference Materials. Geostand. Geoanalytical Res. 21, 115–144. doi:10.1111/j.1751-908X.1997.tb00538.x

Peccerillo, A., 2005. Plio-Quaternary Volcanism in Italy: Petrology, Geochemistry, Geodynamics. Springer-Verlag, Berlin/Heidelberg. doi:10.1007/3-540-29092-3

Peccerillo, A., Frezzotti, M.L., 2015. Magmatism, mantle evolution and geodynamics at the converging plate margins of Italy. J. Geol. Soc. London. 172, 407–427. doi:10.1144/jgs2014-085

Pedregosa, F., Varoquaux, G.G., Gramfort, A., Michel, V., Thirion, B., Grisel, O., Blondel, M., Prettenhofer, P., Weiss, R., Dubourg, V., Vanderplas, J., Passos, A., Cournapeau, D., Brucher, M., Perrot, M., Duchesnay, É., 2011. Scikit-learn: Machine Learning in Python. J. Mach. Learn. Res. 12, 2825–2830.

Petrelli, M., Morgavi, D., Vetere, F., Perugini, D., 2016. Elemental imaging and petro-volcanological applications of an improved Laser Ablation Inductively Coupled Quadrupole Plasma Mass Spectrometry. doi:10.2451/2015PM0465

Petrelli, M., Perugini, D., 2016. Solving petrological problems through machine learning: the study case of tectonic discrimination using geochemical and isotopic data 171. doi:10.1007/s00410-016-1292-2

Petrelli, M., Perugini, D., Alagna, K.E., Poli, G., Peccerillo, A., 2008. Spatially resolved and bulk trace element analysis by laser ablation - Inductively coupled plasma - Mass spectrometry (LA-ICP-MS). Period. di Mineral. 77, 3–21. doi:10.2451/2008PM0001

Petrelli, M., Perugini, D., Moroni, B., Poli, G., 2003. Determination of travertine provenance from ancient buildings using self-organizing maps and fuzzy logic. Appl. Artif. Intell. 17, 885–900. doi:10.1080/713827251

Petrelli, M., Perugini, D., Poli, G., Peccerillo, A., 2007. Graphite electrode lithium tetraborate fusion for trace element determination in bulk geological samples by laser ablation ICP-MS. Microchim. Acta 158, 275–282. doi:10.1007/s00604-006-0731-6

Petrelli, M., Poli, G., Perugini, D., Peccerillo, A., 2005. PetroGraph: A new software to visualize, model, and present geochemical data in igneous petrology. Geochemistry, Geophys.

Geosystems 6, n/a-n/a. doi:10.1029/2005GC000932

Pouget, S., Bursik, M., Cortés, J.A., Hayward, C., 2014. Use of principal component analysis for identification of Rockland and Trego Hot Springs tephras in the Hat Creek Graben, northeastern California, USA. Quat. Res. 81, 125–137. doi:10.1016/j.yqres.2013.10.012

Preece, S.J., Westgate, J.A., Alloway, B. V, Milner, M.W., 2011. Characterization, identity, distribution, and source of late Cenozoic tephra beds in the Klondike district of the Yukon, Canada. Can. J. Earth Sci.

Provost, F., Kohavi, R., 1998. Guest editors' introduction: On applied research in machine learning. Mach. Learn. 30, 127–132. doi:10.1023/A:1007442505281

Raffi, I., 2002. Revision of the early-middle pleistocene calcareous nannofossil biochronology (1.75–0.85 Ma). Mar. Micropaleontol. 45, 25–55. doi:10.1016/S0377-8398(01)00044-5

Rio, D., Raffi, I., Villa, G., 1990. 32. Pliocene-Pleistocene Calcareous Nannofossil Distribution Patterns in the Western Mediterranean. Proc. Ocean Drill. Progr. Sci. Results 107.

Rocholl, A., 1998. Major and Trace Element Composition and Homogeneity of Microbeam Reference Material: Basalt Glass USGS BCR-2G. Geostand. Geoanalytical Res. 22, 33–45. doi:10.1111/j.1751-908X.1998.tb00543.x

Satow, C., Tomlinson, E.L., Grant, K.M., Albert, P.G., Smith, V.C., Manning, C.J., Ottolini, L., Wulf, S., Rohling, E.J., Lowe, J.J., Blockley, S.P.E., Menzies, M.A., 2015. A new contribution to the Late Quaternary tephrostratigraphy of the Mediterranean: Aegean Sea core LC21. Quat. Sci. Rev. 117, 96–112. doi:10.1016/j.quascirev.2015.04.005

Schmid, P., Peltz, C., Hammer, V.M.F., Halwax, E., Ntaflos, T., Nagl, P., Bichler, M., 2000. Separation and Analysis of Theran Volcanic Glass by INAA, XRF and EPMA. Microchim. Acta 133, 143–149. doi:10.1007/s006040070084

Scholkopf, B., Burges, C.J.C., Girosi, F., Niyogi, P., Poggio, T., Vapnik, V., 1997. Comparing support vector machines with Gaussian kernels to radial basis function classifiers. IEEE Trans. Signal Process. 45, 2758–2765. doi:10.1109/78.650102

Shane, P.A.R., Froggatt, P.C., 1994. Discriminant Function Analysis of Glass Chemistry of New Zealand and North American Tephra Deposits 41, 70–81. doi:10.1006/qres.1994.1008

Shi, F., Liu, Y.-Y., Sun, G.-L., Li, P.-Y., Lei, Y.-M., Wang, J., 2015. A support vector machine for spectral classification of emission-line galaxies from the Sloan Digital Sky Survey. Mon. Not. R. Astron. Soc. 453, 122–127. doi:10.1093/mnras/stv1617

Snow, C.A., 2006. A reevaluation of tectonic discrimination diagrams and a new probabilistic approach using large geochemical databases: Moving beyond binary and ternary plots. J. Geophys. Res. 111, B06206. doi:10.1029/2005JB003799

Templ, M., Filzmoser, P., Reimann, C., 2008. Cluster analysis applied to regional geochemical data: Problems and possibilities. Appl. Geochemistry 23, 2198–2213. doi:10.1016/j.apgeochem.2008.03.004

Tomlinson, E., Smith, V., Albert, P., 2015. The major and trace element glass compositions of the productive Mediterranean volcanic sources: tools for correlating distal tephra layers in and around Europe. Quat. Sci. Rev. 118, 48–66. doi:10.1016/j.quascirev.2014.10.028

Ustuner, M., 2015. Application of Support Vector Machines for Landuse Classification Using High-Resolution RapidEye Images: A Sensitivity Analysis. Eur. J. Remote Sens. 48, 403. doi:10.5721/EuJRS20154823

Verma, S.P., Armstrong-Altrin, J.S., 2016. Geochemical discrimination of siliciclastic sediments from active and passive margin settings. Sediment. Geol. 332, 1–12. doi:10.1016/j.sedgeo.2015.11.011

Zuo, R., Carranza, E.J.M., 2011. Support vector machine: A tool for mapping mineral prospectivity. Comput. Geosci. 37, 1967–1975. doi:10.1016/j.cageo.2010.09.014

**Figure Captions**

**Figure 1 -** (a) Distribution of volcanism in the Central and Southern Italy area. Numbers in brackets are the age of volcanism expressed in Ma; (b) Magnification of the square area outlined in (a). Modified after Peccerillo (2005).

**Figure 2 -** Sedimentological-stratigraphic log for the Caio section. The chronostratigraphic scale is from Cohen and Gibbard (2011), Foraminifer Zones are from Iaccarino and Premoli Silva (2007), Nannofossil Zones are from Backman et al. (2012), Volcanoclastic horizons are from Baldanza et al. (2011), and Lithostratigraphic Units are from Baldanza et al. (2014).

**Figure 3 -** (a) Handpicked pyroxenes showing idiomorphic habitus; (b) Secondary Electron Microprobe (SEM) Back Scattered Electrons (BSE) image showing the petrography of Caio pumices; (c-d) melt inclusions in pyroxene.

**Figure 4 -** Classification Scores of the SVM algorithm reported as a function of the $C$ and $\gamma$ parameters.

**Figure 5 -** Confusion Matrix for the application of the SVM algorithm on the different volcanic sources of the Pliocene and Pleistocene Italian magmatism.

**Figure 6 -** (a) Condrite normalized (Nakamura, 1974) REE patterns of Caio tephra samples; (b) Incompatible element patterns normalized to the primordial mantle (McDonough and Sun, 1995) composition for Caio tephra samples; diagrams have been constructed using the Petrograph software (Petrelli et al., 2005).

**Figure 7** – Total Alkali vs. Silica binary diagram for Caio tephra and the different volcanic sources of the recent Italian magmatism (displayed as areas with different colours). The diagram has been constructed using samples from the GEOROC dataset.

**Figure 8** - Nb vs. Zr binary diagram highlighting large overlapping among the different volcanic sources of the Pliocene and Pleistocene Italian magmatism. The diagram has been constructed using samples from the GEOROC dataset. Symbols and colours are the same of Figure 7.

**Table captions**

**Table 1** – Ages and compositional characteristics of Italian magmatic provinces. Data are from Peccerillo, 2005 and Peccerillo and Frezzotti, 2015.

**Table 2** - Major and trace elements analyses of Caio melt inclusions hosted in pyroxene phases.

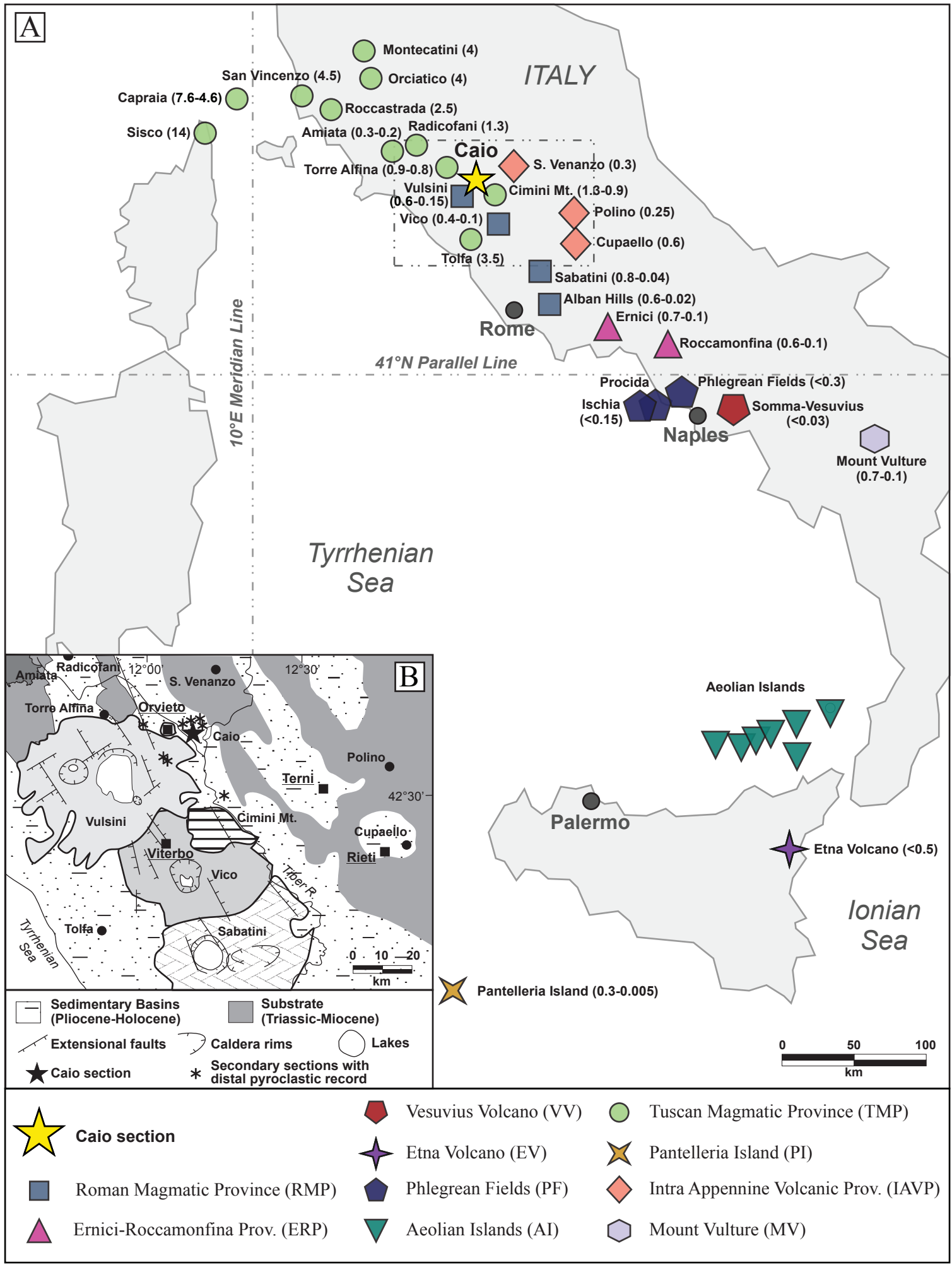

Figure 1

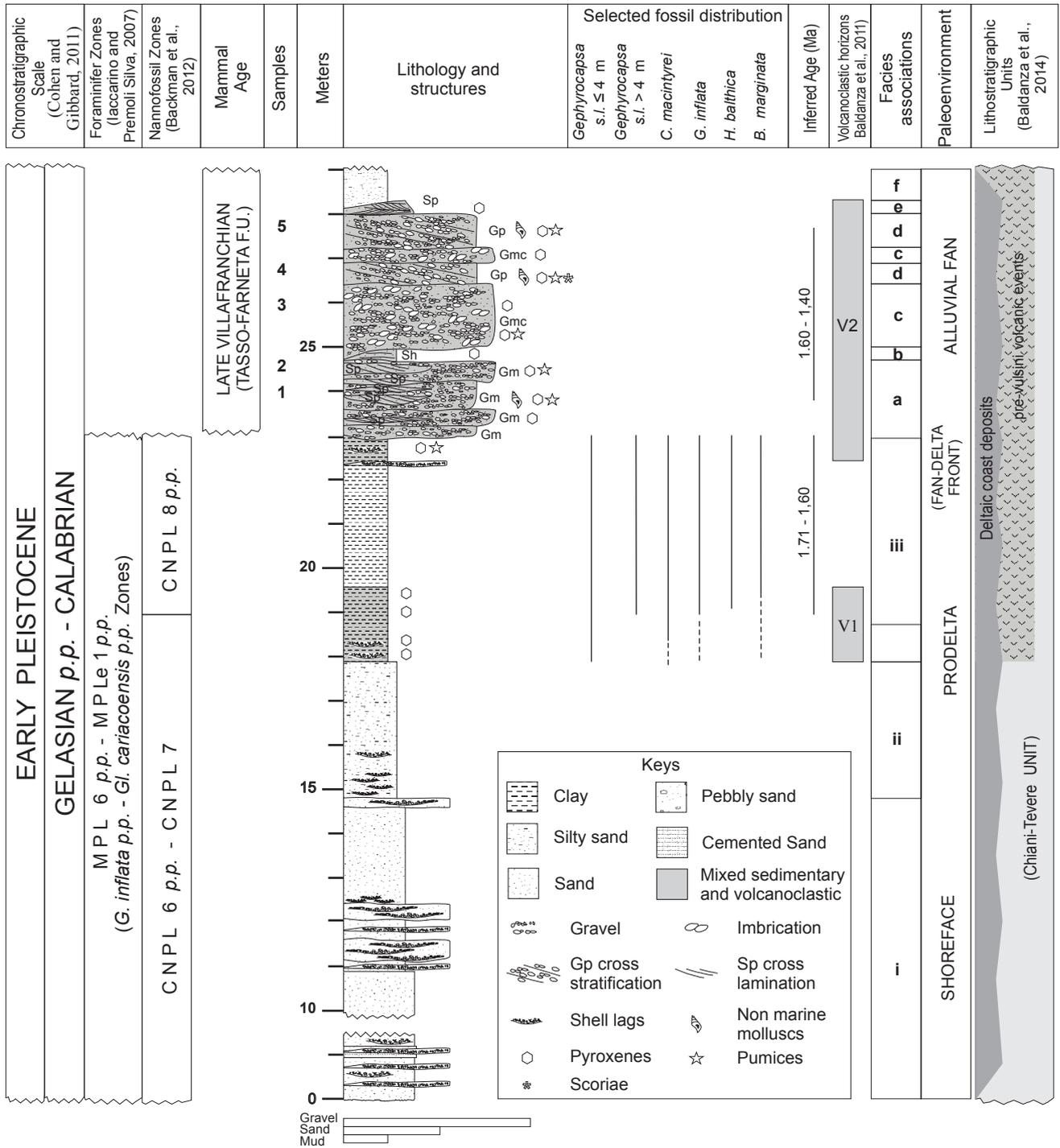

Figure 2

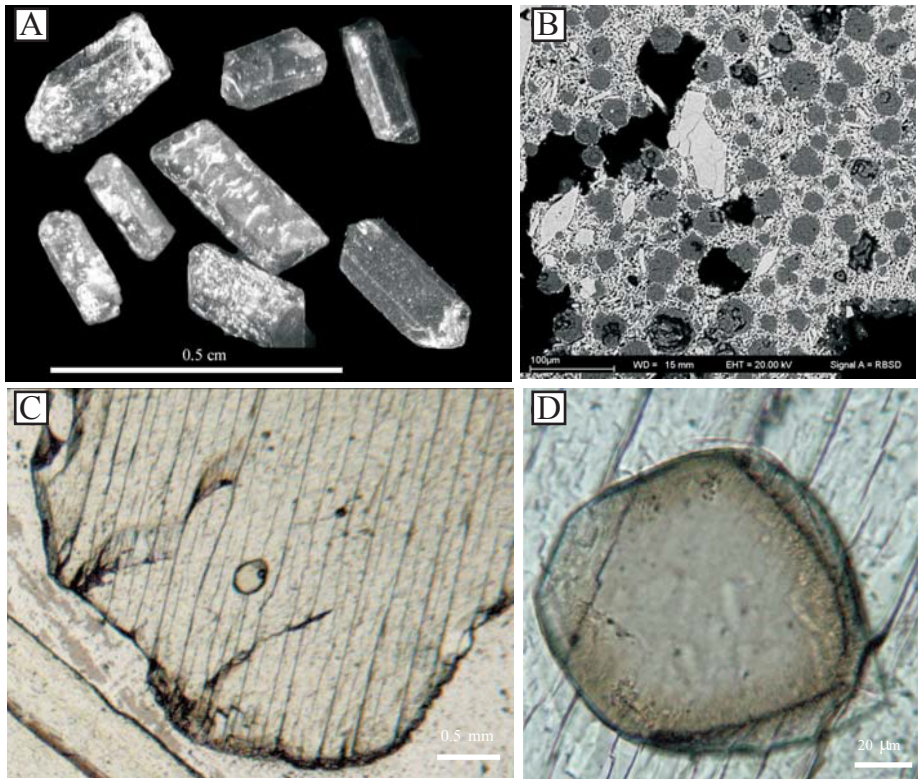

Figure 3

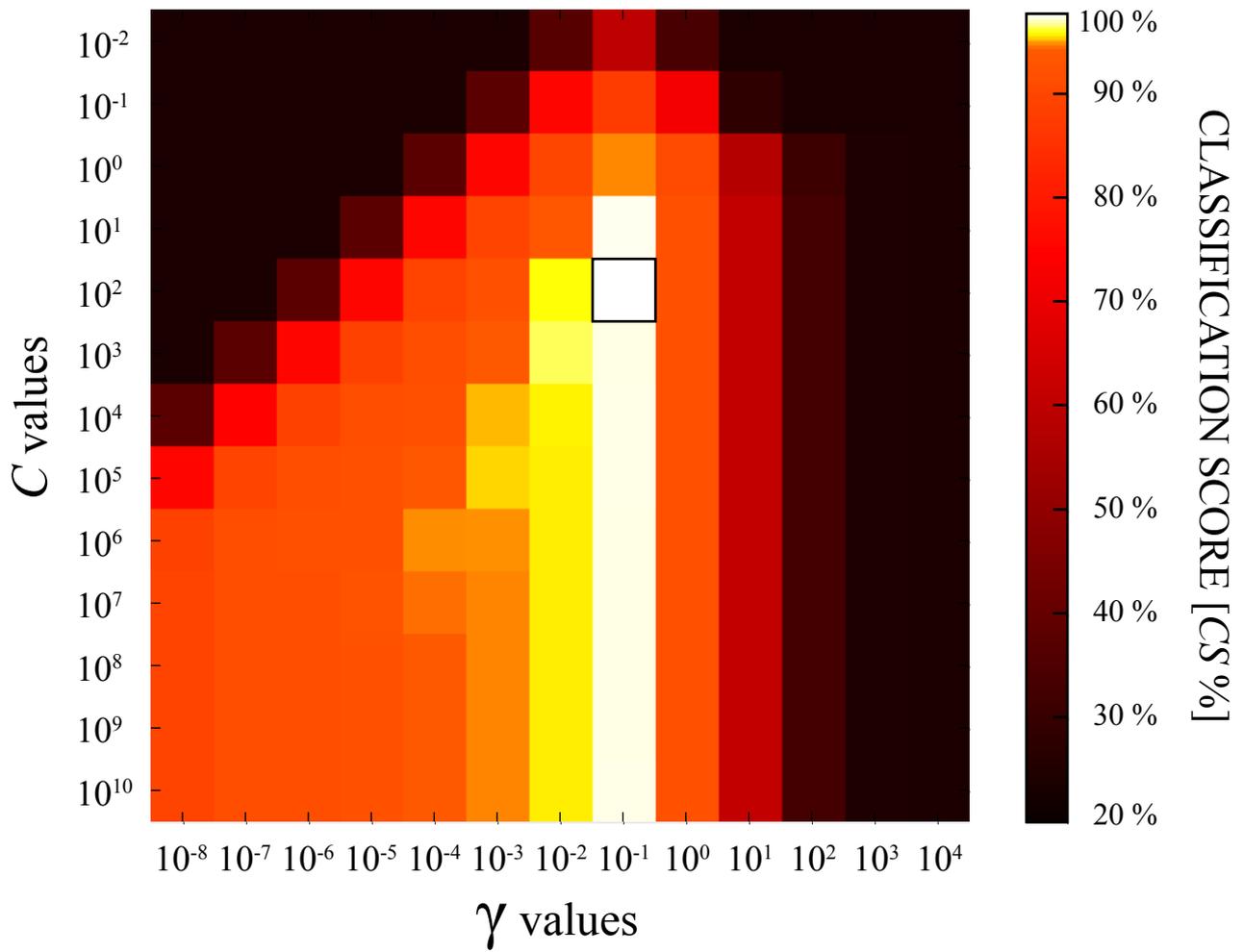

Figure 4

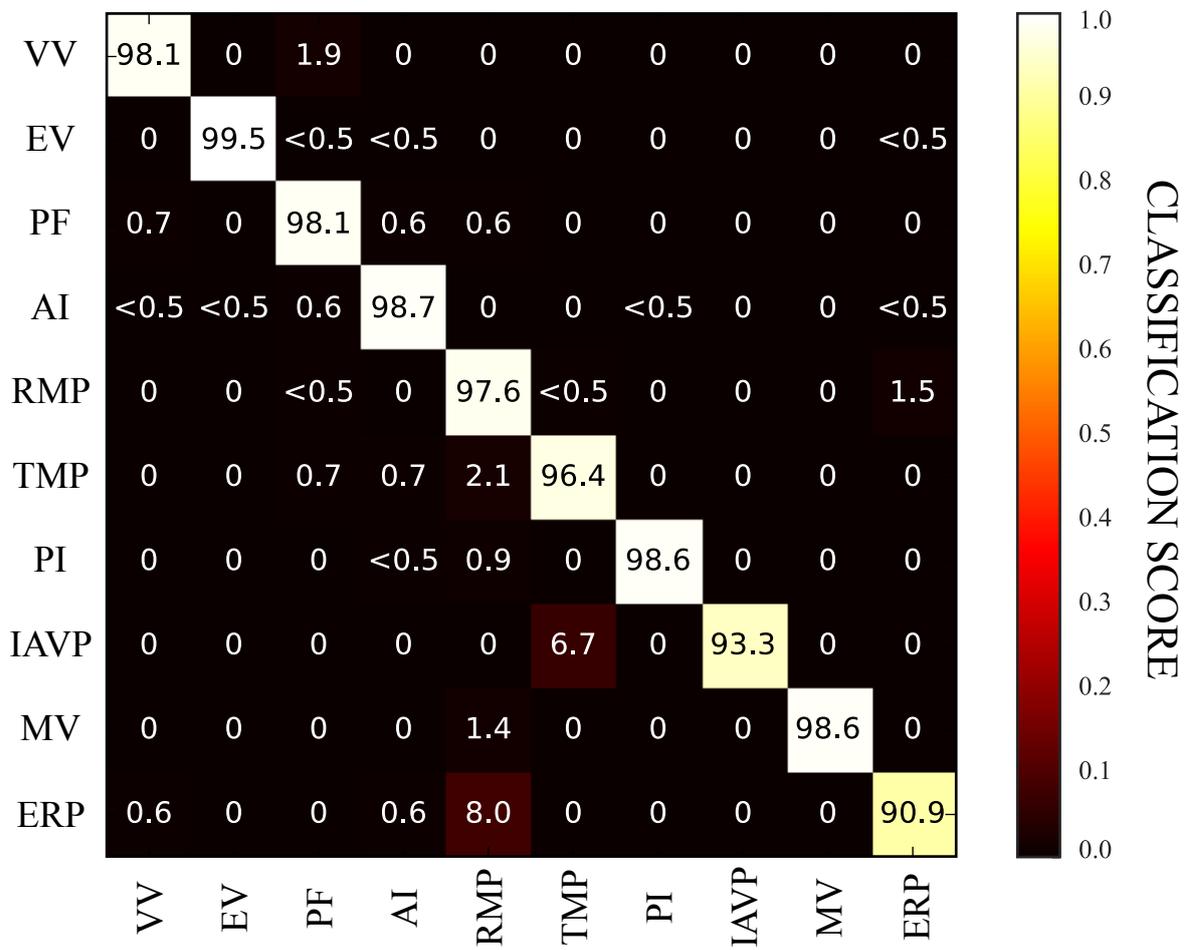

Figure 5

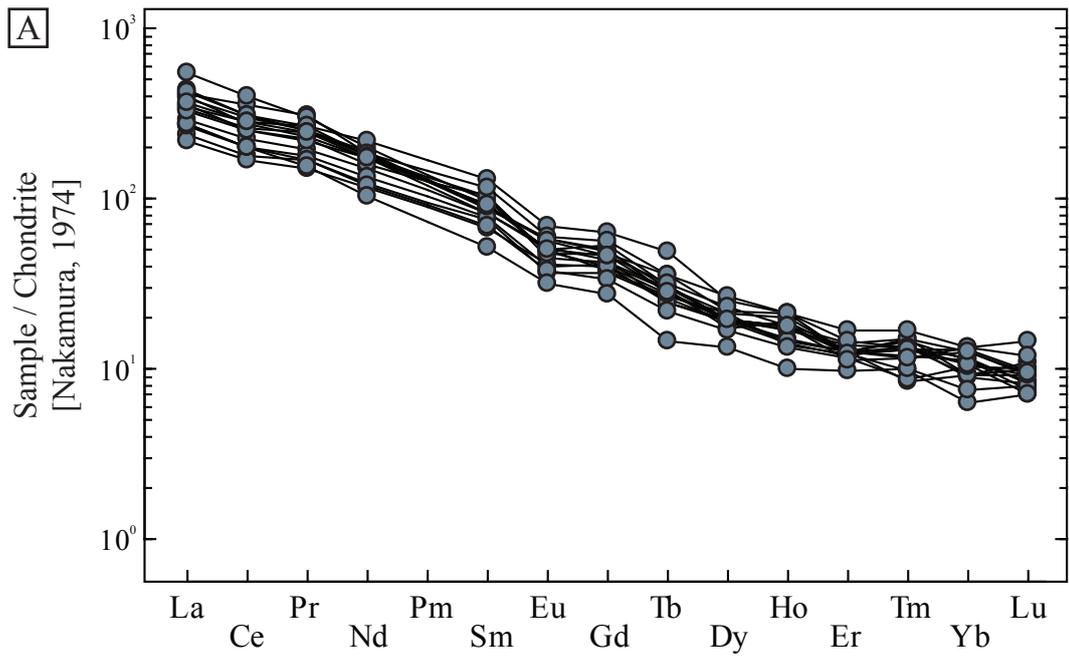
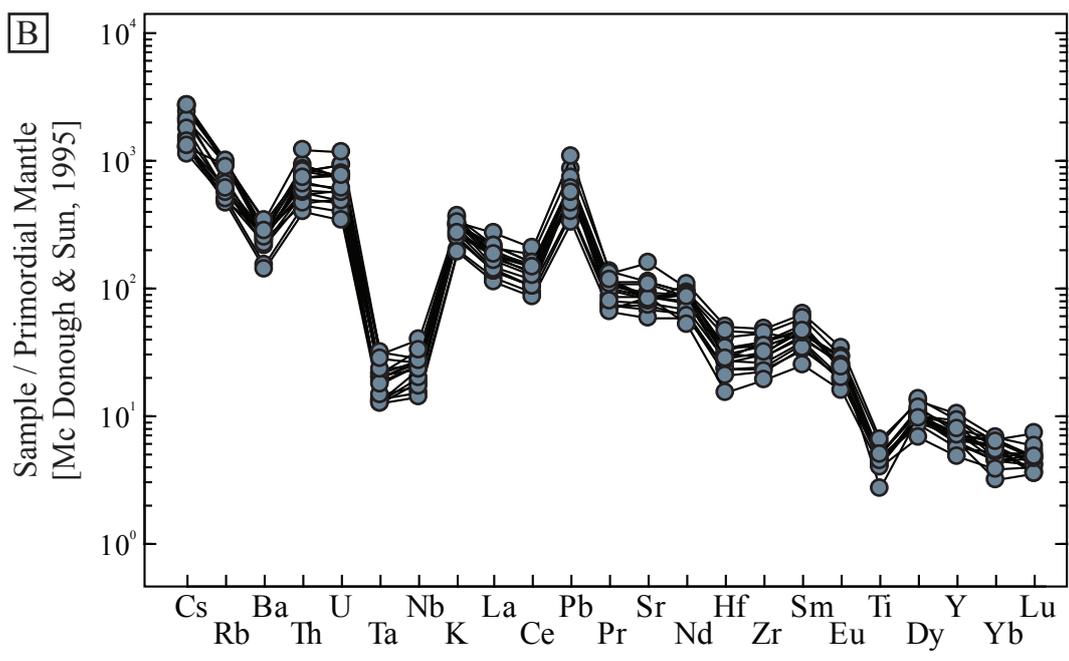

Figure 6

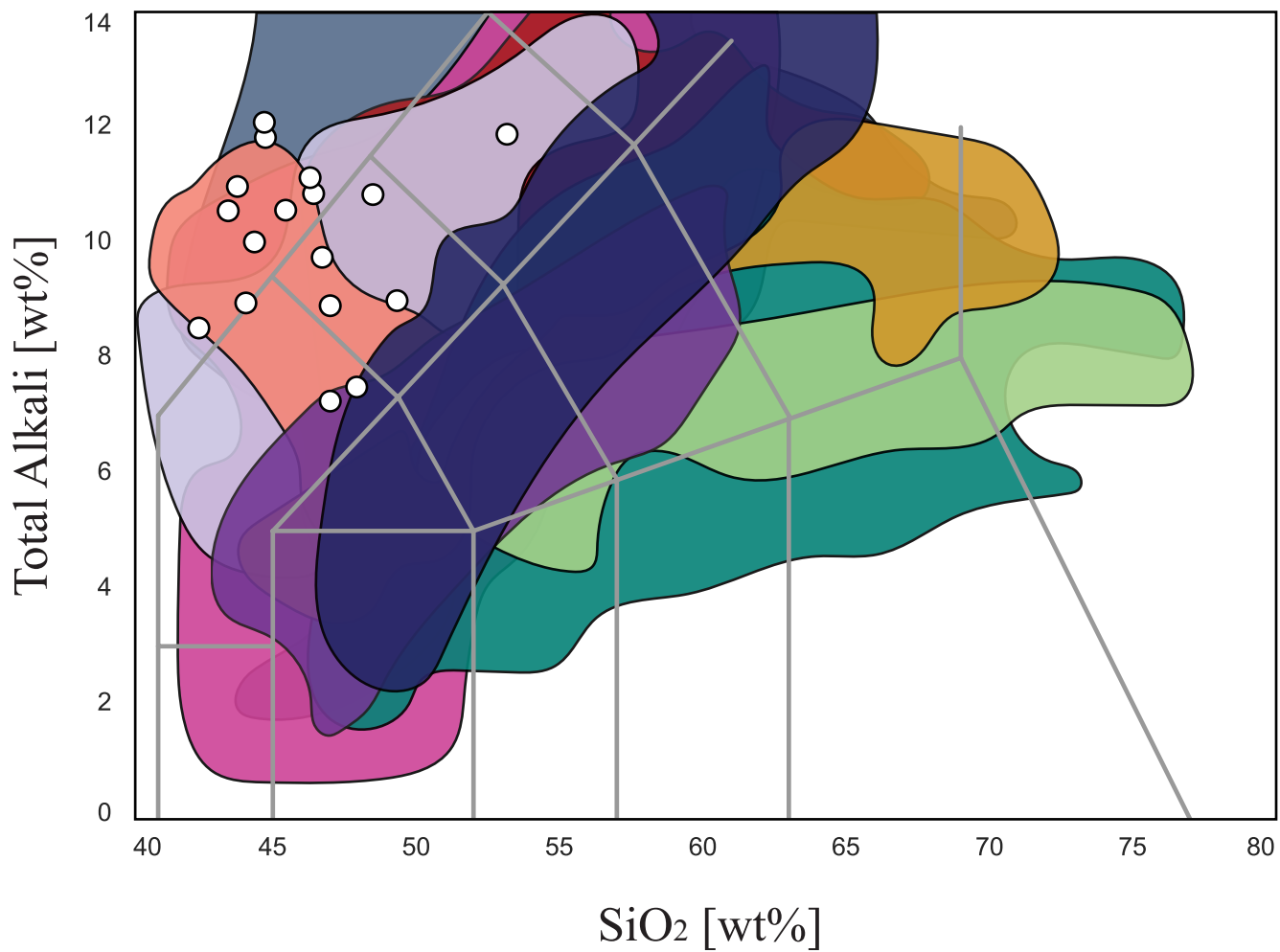

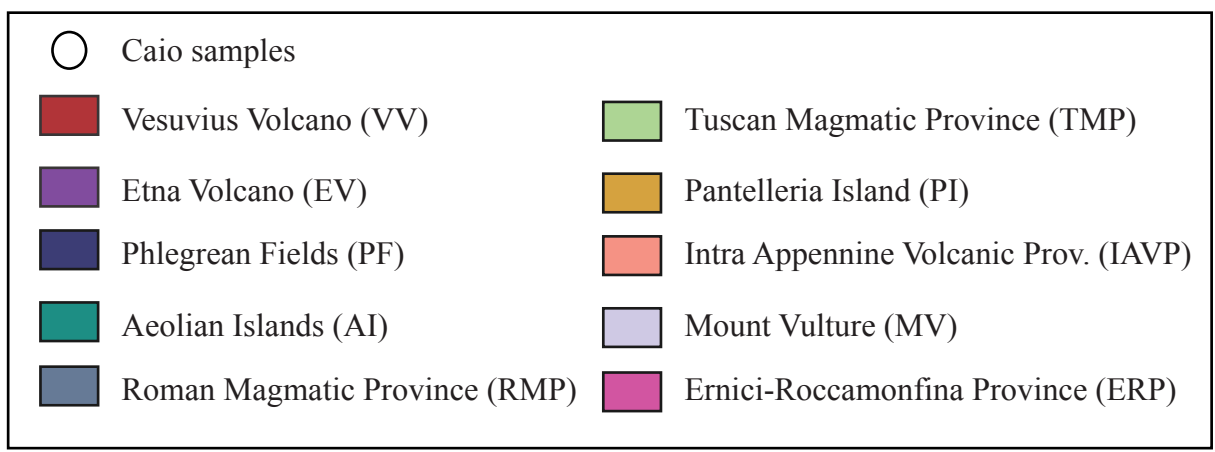

Figure 7

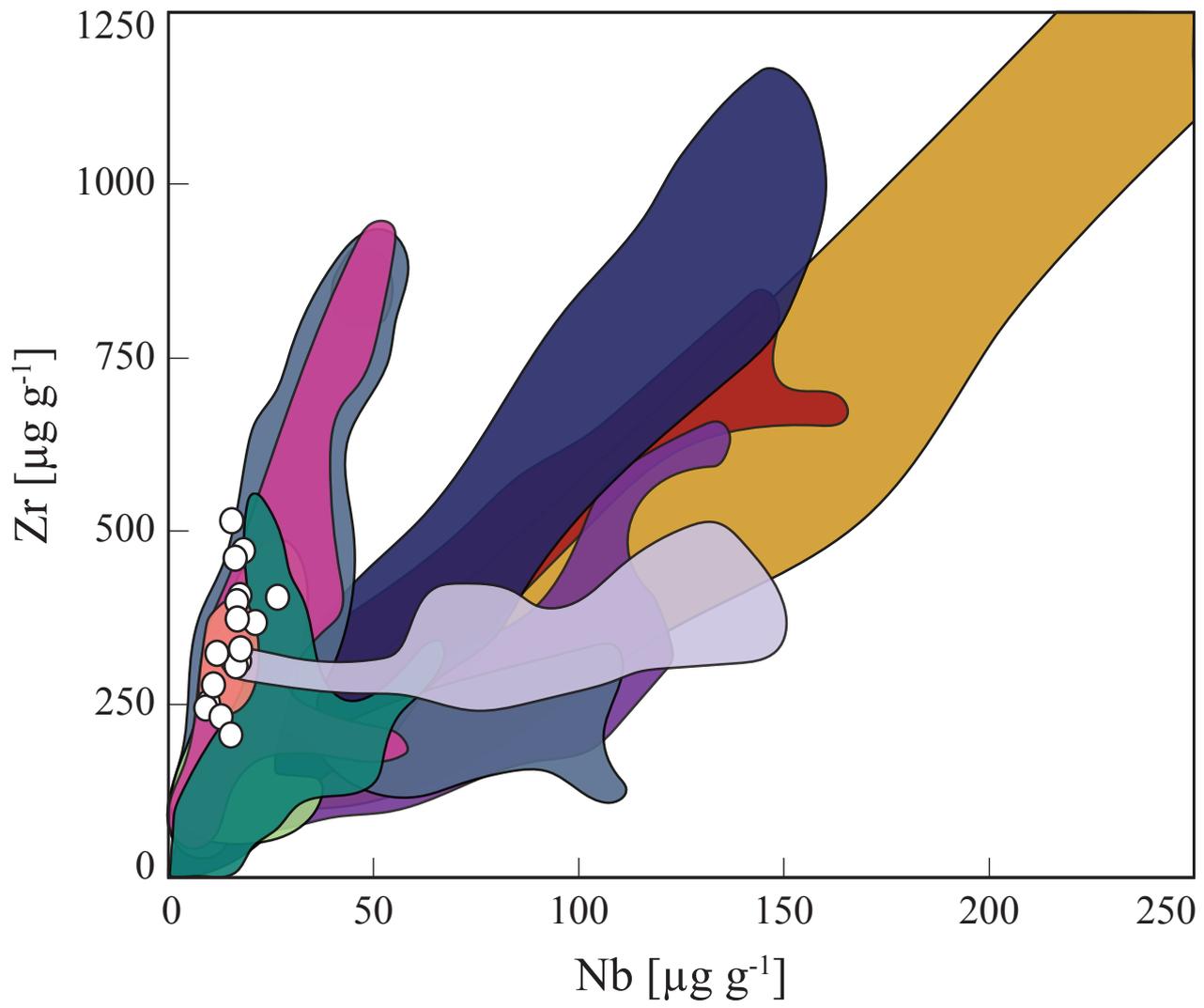

Figure 8

| Magmatic Province | Abbreviation | Age | Petrologic Features of Mafic Rocks |
| --- | --- | --- | --- |
| Tuscan Magmatic Province | TMP | 8.5 Ma - 0.3 Ma | from calc-alkaline to potassic and ultrapotassic affinity |
| Intra Appennine Volcanic Province | IAVP | 0.6 Ma - 0.3 Ma | undersaturated in silica with potassic affinity |
| Roman Magmatic Province | RMP | 0.8 Ma - 0.006 Ma | mostly of ultrapotassic affinity and undersaturated in silica |
| Ernici-Roccamonfina Province | ERP | 0.7 Ma - 0.1 Ma | mostly undersaturated in silica with variable contents in potassium |
| Vesuvius Volcano | VV | 30 ka - 1944 AD | different degrees of silica undersaturation and enrichment in potassium |
| Phlegrean Fields | PF | 0.3 Ma - 1538 AD | slightly undersaturated to oversaturated in silica, different alcali enrichment |
| Mount Vulture | MV | 0.7 Ma - 0.1 Ma | midly to strongly undersaturated in silica, Na-K rich |
| Aeolian Islands | AI | 0.27 Ma - present | mainly calc-alkaline |
| Etna Volcano | EV | 0.6 Ma - present | mainly tholeiitic to Na-alkaline affinity |
| Pantelleria Island | PI | 0.3 Ma - 10 ka | mafic rocks show transitional to alkaline affinity |

| Sample | SiO2 | TiO2 | Al2O3 | Fe2O3t | MnO | MgO | CaO | Na2O | K2O | P2O5 | V | Ga | Rb | Sr | Y | Zr | Nb | Cs | Ba |
|---|---|---|---|---|---|---|---|---|---|---|---|---|---|---|---|---|---|---|---|
| | wt% | wt% | wt% | wt% | wt% | wt% | wt% | wt% | wt% | wt% | µg g$^{-1}$ | µg g$^{-1}$ | µg g$^{-1}$ | µg g$^{-1}$ | µg g$^{-1}$ | µg g$^{-1}$ | µg g$^{-1}$ | µg g$^{-1}$ | µg g$^{-1}$ |
| CAIO-3A7-I1 | 45.7 | 0.8 | 17.4 | 8.7 | 0.1 | 5.0 | 11.1 | 1.4 | 9.2 | 0.5 | 286 | 21 | 571 | 2275 | 29 | 368 | 21.3 | 52 | 2182 |
| CAIO-3A8-I1 | 42.6 | 0.8 | 16.5 | 9.9 | 0.2 | 5.9 | 15.0 | 1.1 | 7.5 | 0.5 | 342 | 15 | 422 | 2212 | 29 | 312 | 17.4 | 42 | 1518 |
| CAIO3B2-I2 | 44.6 | 1.0 | 17.2 | 10.4 | 0.2 | 5.8 | 10.3 | 1.3 | 8.7 | 0.5 | 397 | 17 | 573 | 1719 | 27 | 305 | 16.6 | 57 | 1597 |
| CAIO-3B7-I1 | 46.7 | 0.9 | 22.0 | 7.7 | 0.1 | 3.0 | 8.2 | 2.1 | 8.9 | 0.5 | 210 | 21 | 560 | 1808 | 40 | 472 | 18.4 | 26 | 1940 |
| CAIO-1B2-i1 | 46.9 | 0.8 | 16.1 | 7.5 | 0.2 | 6.5 | 11.7 | 1.5 | 8.3 | 0.6 | 257 | 17 | 398 | 1705 | 32 | 323 | 11.9 | 44 | 1759 |
| CAIO-1B2-i2 | 48.1 | 0.9 | 17.1 | 7.6 | 0.1 | 6.8 | 11.5 | 1.8 | 5.8 | 0.4 | 219 | 16 | 297 | 1332 | 27 | 251 | 9.9 | 27 | 1003 |
| CAIO-1B2-i3 | 47.1 | 0.9 | 17.3 | 7.8 | 0.1 | 7.1 | 12.1 | 1.7 | 5.6 | 0.4 | 224 | 15 | 283 | 1159 | 26 | 244 | 9.2 | 26 | 932 |
| CAIO-1B2-i4 | 47.2 | 1.0 | 19.1 | 7.9 | 0.2 | 5.6 | 9.6 | 1.6 | 7.4 | 0.5 | 234 | 16 | 357 | 1498 | 29 | 277 | 11.1 | 32 | 1397 |
| CAIO-2B2-i1 | 53.5 | 0.6 | 22.9 | 5.1 | 0.2 | 0.8 | 4.9 | 2.5 | 9.5 | 0.2 | 196 | 26 | 563 | 3120 | 30 | 404 | 26.7 | 43 | 2236 |
| CAIO-4A9-i1 | 46.5 | 0.9 | 19.5 | 8.6 | 0.2 | 4.7 | 7.8 | 1.9 | 9.3 | 0.6 | 368 | 22 | 545 | 1694 | 37 | 407 | 17.5 | 56 | 1807 |
| CAIO-4A9-i3 | 45.0 | 1.0 | 19.3 | 9.2 | 0.2 | 5.0 | 8.0 | 1.7 | 10.2 | 0.5 | 366 | 23 | 590 | 1686 | 34 | 398 | 16.8 | 56 | 1881 |
| CAIO-4A1-i1 | 49.5 | 0.9 | 19.9 | 6.9 | 0.1 | 4.7 | 8.6 | 2.0 | 7.0 | 0.4 | 265 | 14 | 341 | 1463 | 24 | 231 | 13.0 | 38 | 1588 |
| CAIO-4A1-i4 | 48.7 | 0.8 | 20.3 | 7.2 | 0.1 | 3.7 | 8.0 | 3.1 | 7.7 | 0.4 | 249 | 16 | 293 | 1710 | 20 | 205 | 15.3 | 29 | 1802 |
| CAIO-4A5-i4 | 44.2 | 1.3 | 15.7 | 9.3 | 0.2 | 6.2 | 13.8 | 1.3 | 7.7 | 0.4 | 281 | 21 | 297 | 1598 | 45 | 515 | 15.6 | 23 | 1461 |
| CAIO-4A5-i5 | 43.6 | 1.2 | 15.8 | 8.5 | 0.2 | 6.2 | 13.5 | 1.4 | 9.2 | 0.4 | 292 | 20 | 332 | 1680 | 39 | 461 | 16.4 | 26 | 1716 |
| CAIO-4A5-i8 | 44.9 | 1.0 | 15.6 | 7.6 | 0.1 | 6.0 | 12.2 | 1.4 | 10.8 | 0.5 | 247 | 20 | 364 | 1612 | 32 | 373 | 16.9 | 28 | 1726 |
| CAIO-5A1-i1 | 44.0 | 0.9 | 18.7 | 8.6 | 0.2 | 4.8 | 11.0 | 1.6 | 9.5 | 0.8 | 314 | 19 | 542 | 2109 | 30 | 329 | 17.7 | 57 | 1687 |

| Sample | La | Ce | Pr | Nd | Sm | Eu | Gd | Tb | Dy | Ho | Er | Tm | Yb | Lu | Hf | Ta | Pb | Th | U |
|---|---|---|---|---|---|---|---|---|---|---|---|---|---|---|---|---|---|---|---|
| | µg g$^{-1}$ | µg g$^{-1}$ | µg g$^{-1}$ | µg g$^{-1}$ | µg g$^{-1}$ | µg g$^{-1}$ | µg g$^{-1}$ | µg g$^{-1}$ | µg g$^{-1}$ | µg g$^{-1}$ | µg g$^{-1}$ | µg g$^{-1}$ | µg g$^{-1}$ | µg g$^{-1}$ | µg g$^{-1}$ | µg g$^{-1}$ | µg g$^{-1}$ | µg g$^{-1}$ | µg g$^{-1}$ |
| CAIO-3A7-I1 | 136 | 305 | 34 | 117 | 18 | 4.5 | 14.0 | 1.27 | 6.6 | 1.05 | 2.8 | 0.36 | 2.5 | 0.29 | 8.6 | 0.81 | 131 | 66 | 18.5 |
| CAIO-3A8-I1 | 126 | 257 | 28 | 117 | 18 | 4.5 | 13.3 | 1.41 | 7.3 | 1.40 | 2.8 | 0.44 | 2.5 | 0.25 | 8.6 | 1.05 | 94 | 65 | 19.0 |
| CAIO3B2-I2 | 95 | 192 | 22 | 95 | 17 | 3.9 | 10.5 | 1.24 | 7.3 | 0.96 | 2.8 | 0.29 | 1.4 | 0.24 | 8.7 | 0.67 | 84 | 59 | 16.3 |
| CAIO-3B7-I1 | 131 | 237 | 26 | 109 | 16 | 4.0 | 11.9 | 1.16 | 6.7 | 1.25 | 2.8 | 0.45 | 2.8 | 0.31 | 11.5 | 1.15 | 101 | 75 | 15.0 |
| CAIO-1B2-i1 | 116 | 218 | 25 | 99 | 21 | 3.5 | 11.6 | 1.42 | 6.7 | 1.16 | 3.1 | 0.40 | 2.0 | 0.34 | 7.3 | 0.46 | 104 | 55 | 11.7 |
| CAIO-1B2-i2 | 78 | 155 | 19 | 78 | 15 | 2.8 | 9.9 | 1.20 | 6.5 | 0.98 | 2.6 | 0.41 | 2.0 | 0.28 | 6.5 | 0.51 | 57 | 35 | 8.1 |
| CAIO-1B2-i3 | 71 | 143 | 17 | 73 | 14 | 3.1 | 10.9 | 1.16 | 6.5 | 1.21 | 2.9 | 0.25 | 2.0 | 0.31 | 6.5 | 0.46 | 50 | 32 | 7.1 |
| CAIO-1B2-i4 | 87 | 173 | 20 | 84 | 16 | 3.1 | 11.3 | 1.37 | 6.0 | 1.32 | 2.9 | 0.40 | 2.6 | 0.28 | 7.7 | 0.54 | 61 | 40 | 9.4 |
| CAIO-2B2-i1 | 179 | 345 | 34 | 126 | 20 | 3.8 | 11.0 | 1.27 | 6.7 | 1.02 | 2.8 | 0.35 | 2.9 | 0.49 | 7.8 | 1.11 | 159 | 98 | 24.0 |
| CAIO-4A9-i1 | 141 | 267 | 29 | 111 | 19 | 3.9 | 14.6 | 1.45 | 6.7 | 1.26 | 3.2 | 0.45 | 2.0 | 0.37 | 9.5 | 0.74 | 102 | 69 | 15.7 |
| CAIO-4A9-i3 | 140 | 263 | 28 | 110 | 21 | 3.7 | 13.1 | 1.69 | 7.5 | 1.46 | 2.7 | 0.39 | 2.9 | 0.33 | 9.3 | 0.68 | 103 | 65 | 14.7 |
| CAIO-4A1-i1 | 89 | 173 | 19 | 76 | 14 | 2.9 | 9.1 | 1.01 | 5.7 | 0.94 | 2.6 | 0.26 | 2.3 | 0.34 | 6.0 | 0.48 | 83 | 37 | 10.3 |
| CAIO-4A1-i4 | 92 | 175 | 17 | 65 | 10 | 2.4 | 7.6 | 0.68 | 4.5 | 0.71 | 2.2 | 0.30 | 1.7 | 0.27 | 4.3 | 0.48 | 109 | 43 | 11.7 |
| CAIO-4A5-i4 | 115 | 246 | 30 | 136 | 26 | 5.2 | 17.6 | 2.28 | 8.7 | 1.50 | 3.7 | 0.51 | 3.0 | 0.40 | 14.2 | 0.77 | 71 | 44 | 9.9 |
| CAIO-4A5-i5 | 110 | 223 | 27 | 114 | 24 | 4.6 | 15.6 | 1.68 | 9.1 | 1.46 | 3.3 | 0.40 | 2.4 | 0.24 | 13.2 | 0.85 | 80 | 47 | 11.5 |
| CAIO-4A5-i8 | 108 | 215 | 25 | 109 | 18 | 4.3 | 12.7 | 1.49 | 7.9 | 1.25 | 2.7 | 0.39 | 2.4 | 0.31 | 9.5 | 0.74 | 83 | 52 | 12.5 |
| CAIO-5A1-i1 | 119 | 243 | 28 | 108 | 19 | 3.9 | 12.7 | 1.34 | 6.6 | 1.25 | 2.5 | 0.35 | 2.8 | 0.32 | 8.0 | 0.79 | 93 | 59 | 15.4 |